\documentstyle[multicol,aps,epsf,tabularx]{revtex}

\begin{document}

\title{Charge Relaxation Resistances and 
Charge Fluctuations in Mesoscopic Conductors}

\author{M.\ B\"uttiker}
\address{Department of Theoretical Physics,
University of Geneva, Geneva, Switzerland}

\date{\today}
\maketitle

\begin{abstract}
A brief overview is presented of recent work which investigates the 
time-dependent relaxation of charge and its spontaneous 
fluctuations on mesoscopic conductors in the proximity of gates. 
The leading terms of the low frequency conductance are determined 
by a capacitive or inductive emittance and a dissipative charge relaxation 
resistance. The charge relaxation resistance is determined by 
the ratio of the mean square dwell time of the carriers in the conductor
and the square of the mean dwell time. The contribution of each 
scattering channel is proportional to {\it half} a resistance quantum. 
We discuss the charge relaxation resistance for mesoscopic capacitors, 
quantum point contacts, chaotic cavities, ballistic wires and for
transport along edge channels in the quantized Hall regime. 
At equilibrium the charge relaxation resistance also determines  
via the fluctuation-dissipation theorem the spontaneous fluctuations 
of charge on the conductor. Of particular interest are the charge
fluctuations in the presence of transport in a regime where the 
conductor exhibits shot noise. At low frequencies and voltages 
charge relaxation is determined by a nonequilibrium charge relaxation 
resistance.

\end{abstract}


\begin{multicols}{2}

\section{Introduction}

The dynamics of charge relaxation and the fluctuations of 
charge represent a basic aspect of electrical conduction 
theory. In this work we are primarily concerned with a 
novel transport coefficient which governs the relaxation 
of excess charge on a mesoscopic conductor towards 
its equilibrium value and also the time-dependent 
spontaneous fluctuations of this charge \cite{btp93}.  
In a macroscopic 
conductor the relaxation of excess charge is determined 
by an $RC$-time constant. Macroscopically, 
electric fields are screened at the surface of conductors
and as a consequence the capacitance $C$
is essentially determined by the geometrical configuration of 
the conductor. In macroscopic circuits, the resistance 
is typically taken to be a dc-resistance which can be found 
by considering a time-independent transport problem. 
Interestingly, for mesoscopic conductors the capacitance
can depend in a significant way on the properties of the mesoscopic 
conductor and its nearby gates. Even more dramatically, the charge 
relaxation resistance can not be found by considering a time-independent 
conduction problem. It is the purpose of this work to present 
a brief review of such charge relaxation resistances.
In particular, we consider the charge relaxation resistance of 
a mesoscopic capacitor, of a quantum point contact, of a 
chaotic cavity coupled to single channel leads or to wide 
quantum point contacts, of a ballistic wire, and of a Hall bar. 
The charge dynamics is of interest in itself but at low temperature
it is also expected to be important to understand dephasing.

At equilibrium the charge on a mesoscopic conductor fluctuates due to 
thermal agitation and due to zero-point fluctuations in the 
occupation numbers of the reservoir quantum channels. The low 
frequency spectrum of the charge fluctuations is precisely 
related to the charge relaxation resistance.
Furthermore, it is interesting to 
ask about the charge fluctuations on a mesoscopic conductor
not only in its equilibrium state but also 
in the presence of transport. At low voltages such a sample 
exhibits shot noise due to the granularity of the electric
charge. We find that the charge fluctuations in this case are again 
determined by a non-equilibrium charge relaxation resistance 
which is a close relative of the equilibrium charge relaxation resistance. 

The discussion of charge relaxation resistances is a special 
topic of a much wider field: the characterization of the dynamic 
and non-linear behavior of mesoscopic samples. We cite here 
only a review \cite{cura}, a few original works\cite{cbprl,cbepl}
and some of the recent 
works \cite{song,ma,sablikov,anat,aron,kuro,photo,agua,tang}
to indicate the breadth of interest in this 
field.

\section{The mesoscopic capacitor} 

\begin{figure}
\narrowtext
\epsfysize=9cm
\epsfxsize=7cm
\centerline{\epsffile{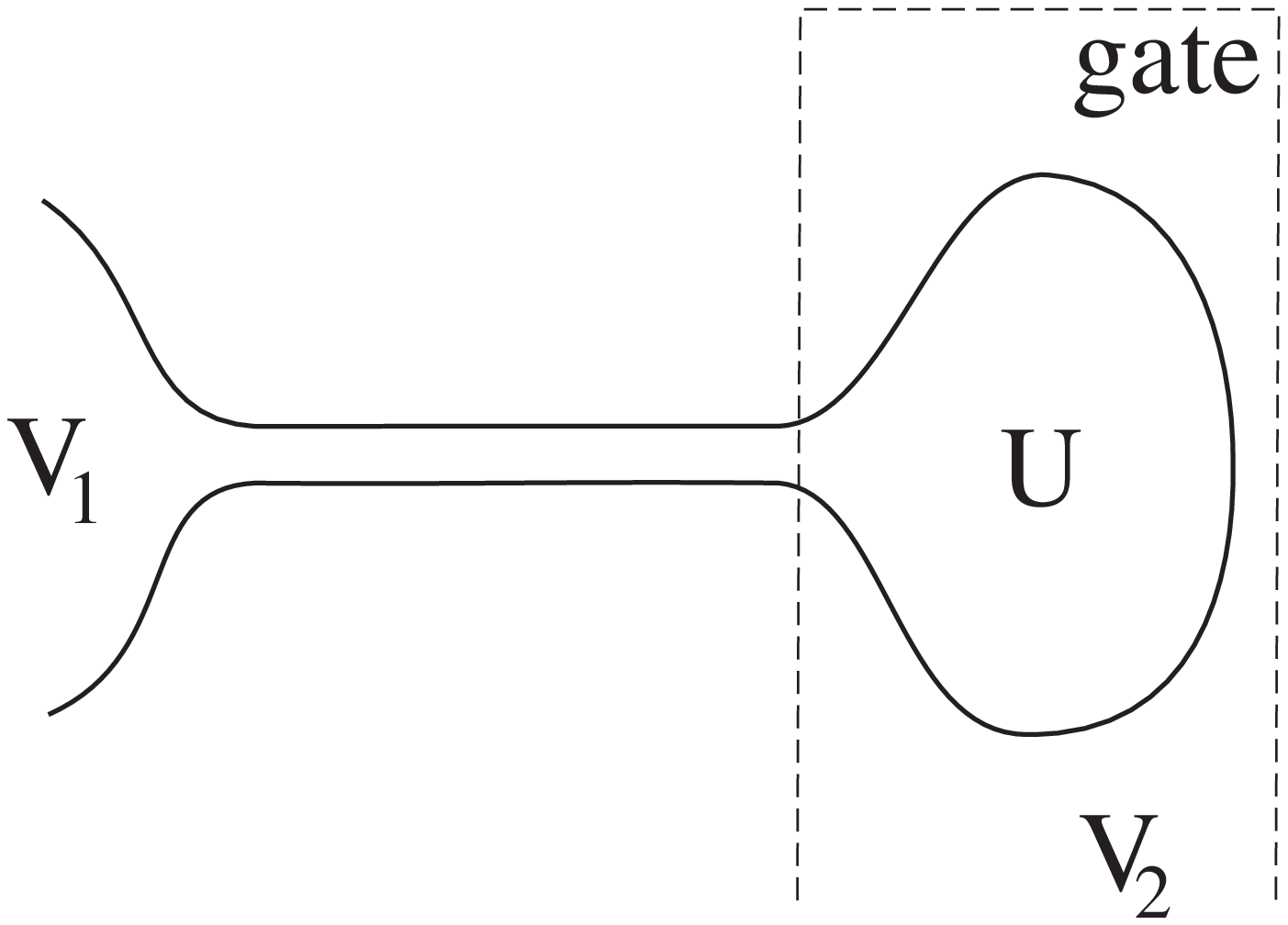}}
\vspace*{0.3cm}
\caption{ \label{mesocap}
Mesoscopic capacitor connected via a single lead to an electron reservoir 
and capacitivly coupled to a gate. $V_1$ and $V_2$ are the potentials applied
to the contacts, $U$ is the electrostatic potential of the cavity. }
\end{figure}

Consider the conductor shown in Fig.~\ref{mesocap}. It consists 
of a small cavity which is via a lead connected to a reservoir
at the potential $V_{1}$ and is capacitively 
coupled to a back gate at the potential
$V_{2}$. Throughout this 
work we treat the back gate as a macroscopic conductor. 
We are interested in the current $I(t)$ driven through this
structure in response to a voltage oscillation  
$V(t) = V_{1}(t) - V_{2}(t)$ where 
$V_{1}(t)$ and  $V_{2}(t)$ are the potentials at the contacts 
$1$ and $2$. We are concerned with linear response and 
without loss of generality can assume that the voltage 
is of the form $V(t) = V(\omega) exp(-i\omega t)$. 
In a macroscopic description we would assume 
that the dynamic conductance $G(\omega) = I(\omega)/V(\omega)$ 
of such an arrangement is given by a series combination 
of a geometrical capacitance $C$ and a resistor with resistance $R$.
This gives a conductance $G^{-1}(\omega) = (-i\omega C)^{-1} +R$ 
which when expanded in powers of $\omega$ has the leading terms 
\begin{equation}
G(\omega) = -i\omega C + \omega^{2} C^{2} R + O(\omega^{3})
\label{g0}
\end{equation}
The first term describes the non-dissipative capacitive response 
due to the displacement current in the circuit and the second term 
is a dissipative term determined by the capacitance and the 
$RC$-time of the circuit. 

Consider now a conductor that is in the mesoscopic regime.
We consider the zero-temperature limit and investigate phase-coherent
transport. We are interested in the low-frequency conductance 
and are main task is to find the mesoscopic analogues of the transport 
coefficients $C$ and $R$ in Eq. (\ref{g0}). As for dc-transport 
we describe the conductor with the help of a scattering matrix 
${\bf s}$ which relates the amplitudes of the incoming currents 
in the mesoscopic conductor to the amplitudes of the 
out-going currents \cite{cura}. Note that for the conductor in 
Fig. \ref{mesocap}
we have only reflection processes. 
If the potentials applied to the conductors are held fixed 
(constant in time) the scattering matrix depends on the 
energy $E$ of the incoming carriers and depends on the 
equilibrium electrostatic potential $U_{eq}$. 
Thus we write for the 
scattering matrix ${\bf s}(E,U_{eq})$.
The electrostatic potential is in general a complicated 
function of position and is function of the dc-voltage difference 
applied across the capacitor. Thus it is generally a complicated task 
to find the scattering matrix. Here we proceed by assuming that this task
has been accomplished and that  ${\bf s}(E,U_{eq})$ is known. 

In the presence of a time-dependent voltage the 
electrostatic potential will also exhibit oscillations.
For the linear response regime of interest here the 
oscillating potential $dU(t)$ represents a small additional 
contribution to the equilibrium potential and the total 
potential is $U(t) = U_{eq} +dU(t)$. 
To find the admittance for the system of interest, we proceed 
in the following way\cite{btp93}. First we investigate the response
to the oscillating potential at contact $1$ and suppose that the internal 
potential is held fixed at its equilibrium value $U_{eq}$. 
In a second step we determine self-consistently the internal 
oscillating potential $dU(t)$. In a third step we find the 
current response to this internal oscillating potential. 
We now consider the simple case, in which the 
oscillating contribution to the internal potential can be taken 
to be spatially uniform within the cavity. Instead of solving
the full Poisson equation to find the internal potential 
we then ask that the total excess charge $dQ$ on the mesoscopic capacitor
plate is related to the variation of the internal potential 
by a geometrical capacitance coefficient $C$ such that 
\begin{equation}
dQ (\omega) = C  dU(\omega) .
\label{u0}
\end{equation}
Note that a description in terms of geometrical 
capacitances is very often also employed in the discussion of 
Coulomb blockade effects.  
We will not here give additional details 
of the derivation but refer the interested reader to 
Refs. \onlinecite{btp93,cura}. 
In the WKB-limit, we express the transport coefficients of interest 
with the geometrical capacitance $C$ of Eq. (\ref{u0})
and the matrix 
\begin{equation}
{\bf {\cal N}} = \frac{1}{2\pi i} 
{\bf s}^\dagger \frac{d{\bf s}}{dE}.
\label{n0}
\end{equation}
Eq. (\ref{n0}) is the phase-delay matrix of Smith \cite{smith}. 
In the WKB limit\cite{mb83,mb90} the derivative with respect to energy 
of the scattering matrix 
is equal to minus the derivative 
of the scattering matrix with respect to the internal potential $U$,  
$d{\bf s}/{dE} = - d{\bf s}/{d(eU)}$. 
The trace of this matrix is the (total) density of states (at constant 
internal potential) of the mesoscopic structure 
\begin{equation}
{\cal D} = Tr{\bf {\cal N}} = \frac{1}{2\pi i} 
Tr\left({\bf s}^{\dagger} \frac{d{\bf s}}{dE} \right) .
\label{d0}
\end{equation}
Eq. (\ref{d0}) is evaluated at the Fermi energy 
(the electrochemical potential) of the mesoscopic structure. 

The capacitance of the mesoscopic structure is now found to be 
a series combination of the geometrical capacitance 
and the contribution from the density of states\cite{btp93}
\begin{equation}
C_{\mu} = C e^{2}{\cal D}/(C + e^{2}{\cal D}) .
\label{c0}
\end{equation}
The capacitance of the mesoscopic structure is of electrochemical 
nature. It depends in an explicite manner via the density of states 
on the properties of the electrical conductor. 
The ratio of the Coulomb energy $E_c = e^{2}/2C$ and the density of 
states is crucial. If $E_c > {\cal D}^{-1}$, then the charging energy is large 
compared to the mean level spacing $\Delta = {\cal D}^{-1}$ and the electrochemical capacitance
is equal to the geometrical capacitance $C_{\mu} = C$.  
If $E_c < {\cal D}^{-1} = \Delta $
the Coulomb interaction is weak and the electrochemical capacitance 
is dominated by the density of states $C_{\mu} = e^{2}{\cal D}$.

Let us next consider the resistance. With the assumption of 
a uniform electrostatic potential inside the cavity and with 
the Poisson equation replaced by Eq. (\ref{u0}) we find\cite{btp93}
\begin{eqnarray}
    R_q = \frac{h}{2e^2} \frac{\mbox{Tr} 
    \left( \cal{N}\cal{N}^\dagger \right)}
	{[\mbox{Tr}({\cal N})]^{2}} .
	\label{Rq} 
\end{eqnarray}
We call this resistance a {\it charge relaxation}
resistance. First note that this is not a mesoscopic dc-resistance 
which would be expressed in terms of transmission probabilities. 
Instead it is the derivative of the scattering matrix with energy 
which enters. Second, note that it appears with a resistance 
quantum which is only half as large as the resistance quantum 
which determines the plateaus in quantum point 
contacts and in the quantized Hall effect. 

In order to better understand these expressions 
we assume that the scattering matrix has been brought into 
diagonal form. We can perform such unitary transformations 
since the expressions given above depend only on the trace 
of the matrices involved. Since we deal with a scattering 
problem which involves only reflections, the diagonal elements 
of the scattering matrix are of the form
\begin{eqnarray}
    s_{nn} = e^{i\phi_n} .
	\label{sm} 
\end{eqnarray}
Here $\phi_{n}$ is the total phase a carrier accumulates 
from its entrance into the cavity to its exit from the cavity.
$n$ labels the number of eigen channels at the Fermi energy. 
The density of states Eq. (\ref{d0}) now becomes simply the 
sum of all energy derivatives of the phases, 
${\cal D} = (1/2\pi) \sum_{n} d\phi_{n}/dE$. Since $\tau_{n} = \hbar d\phi_{n}/dE$
is the time a carrier in the n-th eigen channel spends in the 
conductor $\tau_{D} = (1/N) \sum_{n=1}^{n=N} \hbar d\phi_{n}/dE$
is the average time carriers dwell in the conductor if there are 
$N$ open quantum channels\cite{mb83,mb90}. Thus the density of states is proportional 
to the number of open channels at the Fermi energy and is 
proportional to the average dwell time, 
${\cal D} = N \tau_D/2\pi \hbar$. If we introduce the Coulomb induced, 
geometrical relaxation time $\tau_{g} = h/e^{2}C$, the  expression
for the electrochemical capacitance
Eq. (\ref{c0}), can equally well be expressed in terms of 
$\tau_{g}$ and the average dwell time. For $\tau_{\mu} = (h/e^{2}C_{\mu})$
we find 
\begin{equation}
\frac{1}{\tau_{\mu}} = \frac{1}{\tau_{g}} + \frac{1}{N\tau_{D}}.
\label{c1}
\end{equation}
In a basis in which the scattering matrix is diagonal 
the charge relaxation resistance is,
\begin{eqnarray}
    R_q &=& \frac{h}{2e^2} \frac{
    \left(\sum_{n} (d\phi_{n}/dE)^{2}\right)}
	{(\sum_{n} (d\phi_{n}/dE))^{2}} .
	\label{Rq1} 
\end{eqnarray}
Thus the charge relaxation resistance is proportional 
to the sum of the squares of the dwell times divided by 
the square of the sum of the dwell times, 
\begin{eqnarray}
    R_q &=& \frac{h}{2e^2} \frac{
    \left(\sum_{n} \tau^{2}_{n}\right)}
	{[\sum_{n} \tau_{n}]^{2}} .
	\label{Rq2} 
\end{eqnarray}
For the case of a single channel, we see immediately that 
the charge relaxation resistance is quantized 
\begin{eqnarray}
    R_q = \frac{h}{2e^2}.
	\label{Rq3} 
\end{eqnarray}
For a spin degenerate channel the quantized resistance would be 
$R_q = {h}/{4e^2}$. The factor of two is significant. 
It is a consequence of the fact that in the mesoscopic conductor 
shown in Fig. \ref{mesocap} there is only one reservoir conductor interface
at which the non-equilibrium carrier distribution 
inside the conductor must relax to the equilibrium distribution of the 
the reservoir \cite{btp93}. We note here only that a doubling of the 
quantum is obtained for interface resistances: If we subtract the 
the Landauer resistance $R = (h/e^{2}) (1-T)/T$ from the two 
terminal resistance $R = (h/e^{2}) 1/T$ the difference is a
resistance $(h/e^{2})$ which can be viewed as the sum of two sample
reservoir-interface resistances $(h/2e^{2})$ as discussed by 
Imry \cite{imry} and Landauer \cite{land}. A closer examination
shows, however, that the interface resistance depends 
on the way voltages are measured and depends on whether or not 
we have phase-coherent transport \cite{butt89,gram}.
Another situation in which a doubling of the conductance quantum 
occurs is the conductance of a hybrid normal and superconducting 
interface and recently it has been argued that also this doubling is 
connected less to the fact that we have Andreev scattering but instead 
to the fact that we have a conductor connected to only one (normal) electron 
reservoir\cite{sols}. Such analogies clearly have their limit: In our case 
the potential is dynamic and not static.  
In general, for an $N$-channel lead connecting the 
mesoscopic capacitor plate and the reservoir, the charge relaxation
resistance is not quantized. We note that if both "plates" of the 
capacitor are mesoscopic, then the charge relaxation resistance 
acquires two contributions, one from the connection of each plate 
to a reservoir\cite{btp93}. In a such a geometry, in the single channel limit,  
the charge relaxation resistance is $R_q = {h}/{e^2}$.

The charge relaxation resistance determines the first dissipative 
term in the expansion of the dynamic conductance in terms of 
frequency. Via the fluctuation dissipation theorem 
it must thus also be related to the spontaneous 
current fluctuations in the capacitor. 
Indeed, for a voltage controlled circuit,
Ref. \cite{btp93}
finds that the noise power spectrum of the equilibrium current fluctuations,
$2\pi \delta(\omega + \omega^{\prime}) S_{II}(\omega) = (1/2)
\langle I(\omega)I(\omega^{\prime})+ I(\omega^{\prime})I(\omega) \rangle$
is in the classical limit $kT > \hbar \omega$ given by 
\begin{eqnarray}
S_{II}(\omega) = 2kT \omega^{2} C^{2}_{\mu}R_q .
	\label{SI} 
\end{eqnarray}
The current fluctuations are accompaned by fluctuations 
in the internal potential $U$. Since  $dI(\omega) = - i\omega dQ (\omega)$
and $dQ (\omega) = C dU (\omega)$ 
we find for the fluctuation spectrum of the internal potential $U$,  
\begin{eqnarray}
S_{UU}(\omega) = 2kT (C_{\mu}/C)^{2} R_q .
	\label{SU} 
\end{eqnarray}
The ratio $C_{\mu}/C$ tends to $1$ in the limit $E_c >>e^{2}{\cal D}$
and tends to $0$ in the limit $E_c << e^{2}{\cal D}$. 
Note that these fluctuations in the internal potential occur 
for a voltage controlled circuit, i. e. the voltage measured at the 
terminals (reservoirs) does not fluctuate. 
For an infinite external impedance the current does 
not fluctuate but the voltage measured at the terminals 
now fluctuates with a low frequency spectrum $S_{VV}(\omega)$. 
The spectrum of external voltage fluctuations 
$S_{VV}(\omega)$ is at low frequencies just determined by the 
charge relaxation resistance alone, 
\begin{eqnarray}
S_{VV}(\omega) = 2kT R_q .
	\label{SV} 
\end{eqnarray}

In the quantum limit and in the low frequency limit of interest here,
$kT$ is replaced by $\hbar |\omega|$ and thus the 
current fluctuation spectrum is third order in frequency and 
the voltage fluctuation spectra are first order in frequency. 

We have emphasized that for the derivation of Eq. (\ref{Rq})
we have described the potential with a single parameter $U$. 
A more general formula for the charge relaxation resistance, valid 
for a general potential landscape $U({\bf r})$ is given in Ref. 
\onlinecite{mbjmp}. Both the nominator and the denominator contain
the Coulomb interaction. In such a more general formulation the universality
of the charge relaxation resistance in the one-channel limit might be 
lost.  Furthermore in our discussion, electron interactions
are treated in the random phase approximation. Especially in the few channel 
limit, even in the presence of open 
leads, charge quantization effects might be important \cite{aleiner}
and these are neglected in our treatment.

\section{Charge Relaxation Resistances of Chaotic Cavities}

To gain further insight into the physical meaning of the charge 
relaxation resistance, we consider in this section the charge relaxation 
resistance of cavities which are 
in the classical limit chaotic \cite{marcus,beenakker}. 
We consider an ensemble 
of cavities which all have 
the same mean level spacing but in which individual members exhibit small 
fluctuations in their shape. To find the distribution of charge 
relaxation resistances we need to know how the eigenvalues
of the Wigner-Smith matrix are distributed. A description 
of the statistical distribution of these times was obtained only 
recently. Fyodorov and Sommers\cite{fyod2} used the supersymmetric 
sigma model to obtain the distribution of the phase-delay time 
for a unitary ensemble of cavities connected to a perfect single channel lead. 
Independently  Gopar et al. \cite{gopar1,gopar2} used a conjecture by Wigner 
to find the distribution of the delay time for the orthogonal, unitary and 
symplectic ensemble again coupled to a single channel perfect lead. 
An important advance was made by Brouwer, Frahm and Beenakker\cite{brouwer2} 
who found the distribution of eigenvalues of the Wigner-Smith 
matrix for a cavity connected to an N-channel lead. These advances 
have permitted to investigate the fluctuations in the capacitance 
coefficient\cite{gopar1,brouwer2} given by Eq. (\ref{c0}). Subsequently, 
these results were used to find the distribution of the transconductance 
of a chaotic cavity\cite{brouwer3}, the thermal conductance\cite{stijn98}
and the charge relaxation resistances\cite{pedersen} 
of chaotic cavities which we now discuss. Before proceeding, we mention only
that delay time fluctuations are also of interest in optical 
wave propagation problems and are a subject of 
experimental interest\cite{genack}. 

\begin{figure}
\narrowtext
\epsfysize=9cm
\epsfxsize=7cm
\centerline{\epsffile{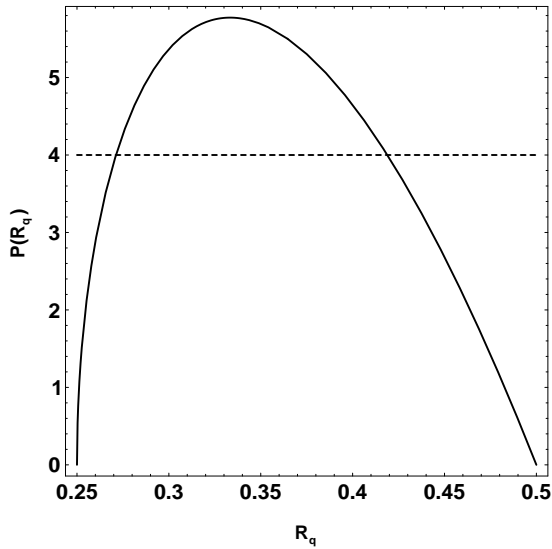}}
\vspace*{0.3cm}
\caption{ \label{qdrq}
Distribution of the charge relaxation resistance of a chaotic quantum dot
for the orthogonal ensemble (dashed line) and the unitary ensemble (solid line) 
line. After Ref. \protect\cite{pedersen}}.
\end{figure}

For a cavity coupled to a single channel lead\cite{gopar1} the charge relaxation 
resistance is according to Eq. (\ref{Rq2}) universal and given by 
half a resistance quantum. It is sharp and exhibits no fluctuations
from one ensemble member to another. 
The situation changes if we consider a cavity connected to two 
perfectly transmitting channels, the case considered by Pedersen, van Langen and 
the author in Ref. \onlinecite{pedersen}. 
For $N = 2$ we see immediately that 

\begin{eqnarray}
    R_q &=& \frac{h}{2e^2} \frac{
    \left(\tau^{2}_{1} + \tau^{2}_{2}\right)}
	{[\tau_{1} + \tau_{2}]^{2}} 
	\label{Rq4} 
\end{eqnarray}
takes the maximum value $h/{2e^2}$ if one of the eigen values 
vanishes and takes the minimum value $h/{4e^2}$ if both eigenvalues 
are identical. A typical sample will thus exhibit a charge relaxation 
resistance between these bounds. Ref. \onlinecite{pedersen}
finds for the distribution function  $P(R_q)$ of the charge 
relaxation resistance 
\begin{equation}
    P(R_q) = \left\{ 
	\begin{array}{ll}
	    4, & \beta=1, \\
	    30(1-2R_q)\sqrt{4R_q-1}, & \beta=2.
	\end{array}
	\right.
	\label{distr}
\end{equation}
Here $\beta =1$ applies to an orthogonal ensemble (no magnetic field)
and $\beta = 2$ applies to a unitary ensemble in which time-reversal
symmetry is broken by a magnetic field. These two distribution functions
are shown in Fig.~\ref{qdrq}.In the orthogonal case the distribution is uniform: 
every charge 
relaxation resistance between $R_q = h/4e^{2}$ and $R_q = h/2e^{2}$
is equally probable. In contrast for a unitary ensemble the limiting
cases in which one of the eigenvalues vanishes or in which both eigenvalues
are equal is not likely and the distribution function is peaked with a 
maximum at $R_q = h/3e^{2}$. 

For cavities connected to a single lead with many quantum channels
it is possible to find the charge relaxation
resistance 
via a $1/N$ expansion. The technique necessary to carry out such 
expansions is discussed in detail in Ref. \onlinecite{brouwer4}. 
The dynamic conductance of a cavity coupled to contacts with a large 
number of channels was investigated by Brouwer and the author\cite{brouwer1}. 
In the large channel limit the distribution is gaussian 
and the charge relaxation is well characterized by its ensemble average 
value. The ensemble averaged charge relaxation resistance is 
\begin{eqnarray}
    R_q &=& \frac{h}{2e^2} \left(\frac{1}{N} 
    + \frac{2-\beta}{\beta} \frac{1}{N^2} + O(\frac{1}{N^3}) \right) .
    \label{RqN} 
\end{eqnarray}
The term proportional to ${1}/{N^2}$ is the weak localization 
correction from enhanced backscattering. Note that 
this resistance is quantized only on the average. 

\section{Charge Relaxation in Two Probe Conductors}

\subsection{Equilibrium and Nonequilibrium Charge Relaxation} 

\begin{figure}
\narrowtext
\epsfysize=9cm
\epsfxsize=7cm
\centerline{\epsffile{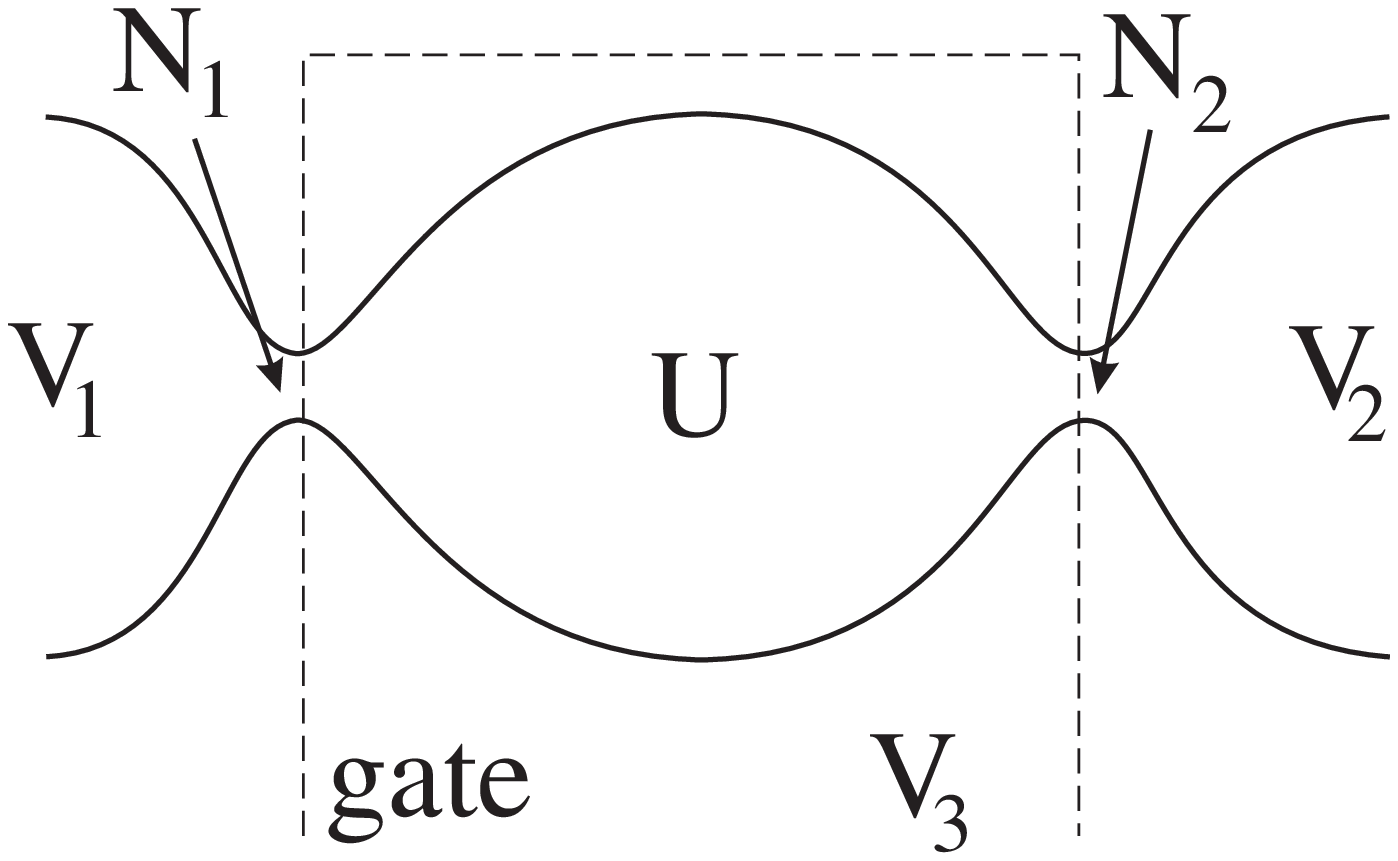}}
\vspace*{0.3cm}
\caption{ \label{twoprobe}
Two probe conductor connected to two electron-reservoirs and 
capacitivly coupled to a gate.}
\end{figure}

Consider now conductors which 
are connected to two or more contacts (see Fig.~\ref{twoprobe}).
In addition the conductor is coupled capacitively
to a gate (or a number of gates) with a (total) geometrical capacitance $C$. 
Let us label the contact to the gate as $3$. 
For the ac-transport problem we have then a conductance problem 
with three contacts. The dynamic conductance $G_{\alpha\beta} (\omega)$
is defined as $dI_{\alpha}(\omega)/dV_{\beta}(\omega) = 
G_{\alpha\beta} (\omega)$ with $\alpha =1, 2, 3,$ 
labelling the contact at which the current is measured 
and $\beta =1, 2, 3$ the contact at which the oscillating potential 
is applied. Thus the dynamic conductance is specified by a three times 
three conductance matrix. 

The two probe conductor acts as a simple mesoscopic capacitor to 
the gate if we consider the case where the voltages at 
the contact of the conductor oscillate in synchronism:
$dV_{1}(\omega) = dV_{2}(\omega)$. The conductance $G_{33}$
has then the same general form as that of a mesoscopic 
capacitor $G_{33}(\omega) = -i\omega C_{\mu,33} + 
\omega^{2} C_{\mu,33} R_{q,33}$
with $C_{33} \equiv C_{\mu}$ the conductor to gate 
capacitance and $R_{q,33} = R_q$, where 
$C_{\mu}$ and $R_q$ are, as we now show, 
given by Eq. (\ref{c0}) and Eq. (\ref{Rq}) obtained above.
The scattering matrix 
which describes the conductor with multiple contacts is specified by matrices 
${\bf s}_{\alpha\beta}$ which give the outgoing current 
amplitudes  at contact $\alpha$ in terms of the incident 
current amplitudes at contact $\beta$. It is useful at this point 
to introduce the matrices\cite{pedersen}
\begin{equation}
{\cal N}_{\delta\gamma} = \frac{1}{2\pi i} \sum_\alpha 
s_{\alpha\delta}^\dagger \frac{ds_{\alpha\gamma}}{dE}. 
\label{d10}
\end{equation}
The diagonal element $\delta = \gamma$ represents the contribution 
to the density of states of carriers injected from contact 
$\gamma$ and is called 
the {\it injectance} of contact $\gamma$. The off-diagonal elements 
start to play a role if we are interested in the charge relaxation resistance 
and in the fluctuations of the charge. 
Thus the total density of states of the conductor is given by 
the diagonal elements of this matrix 
${\cal D} = \sum_{\gamma} \mbox{Tr} ({\cal N}_{\gamma\gamma})$;
where the trace is again over all quantum channels. 
Together with the geometrical 
capacitance the density of states 
determines the capacitance of the sample 
vis-a-vis the gate $C_{\mu,33} \equiv C_{\mu}$
with $C_\mu^{-1} = C^{-1} + ( e^2 {\cal D})^{-1}$, 
i. e. the same result that we have found already for the 
mesoscopic capacitor. 
The charge relaxation resistance is $R_{q,33} \equiv R_q$
with 
\begin{eqnarray}
    R_q &=& \frac{h}{2e^2} \frac{\sum_{\gamma\delta} \mbox{Tr} 
    \left( \cal{N}_{\gamma\delta} \cal{N}^\dagger_{\gamma\delta} \right)}
	{[\sum_{\gamma} \mbox{Tr}({\cal N}_{\gamma\gamma})]^{2}} .
	\label{Rq5} 
\end{eqnarray}
Note, that this is again the same result as Eq. (\ref{Rq}) 
if we understand the matrix ${\bf s}$ to be composed of all 
four matrices ${\bf s}_{\alpha\beta}$. 

Conductors with two or more probes also permit to drive 
a current through the conductor. 
In the zero-frequency and zero-temperature limit,
for a small applied voltage $V$,  
the conductor exhibits shot noise in the transport current.
The spectral density of the current fluctuations due to the shot noise 
can also be 
expressed with the help of the scattering matrix \cite{buttiker90} 
and is given by
$S_{II}(\omega = 0) = (e^{2}/h) Tr({\bf s}^{\dagger}_{11}  
{\bf s}_{11}{\bf s}^{\dagger}_{21}{\bf s}_{21})e|V|$. 
Here we are in particular interested 
in the charge fluctuations on the conductor 
in such a situation. In the zero-temperature, 
low-frequency limit, if current flow is from contact $1$ to contact $2$,
we find for the charge fluctuations 
to leading order in the applied
voltage $V$, 
\begin{equation}
S_{QQ}(\omega) = 2 C_{\mu}^{2} R_{v} |eV|
\label{sqq}
\end{equation}
which defines a non-equilibrium charge relaxation resistance \cite{pedersen} 
\begin{eqnarray}
R_v  = \frac{h}{e^2} \frac{\mbox{Tr} 
    \left( {\cal N}_{21} {\cal N}^\dagger_{21}\right)}
    {[\sum_{\gamma} \mbox{Tr}({\cal N}_{\gamma\gamma})]^{2}}. 
	\label{Rv} 
\end{eqnarray}
Thus the non-equilibrium noise is determined the non-diagonal elements 
of the density of states matrix Eq. (\ref{d10}).

by a non-diagonal
element of the density of states matrix. If both the frequency and 
the voltage are non-vanishing we obtain to leading order
in $\hbar \omega$ and $V$, $S_{QQ}(\omega) = 
2 C_{\mu}^{2} R(\omega,V) \hbar |\omega|$ with a resistance 
\begin{equation}
    R(\omega,V)\hbar|\omega|=\left\{
	\begin{array}{ll}
	    R_q \hbar|\omega| , & \hbar|\omega| \geq e|V| \\
	    R_q \hbar|\omega| + R_V (e|V|-\hbar|\omega|) , &
		\hbar|\omega| \leq e|V|
\end{array} \right. \label{effectiveR}
\end{equation}
which is a frequency and voltage 
dependent series combination of the resistances $R_{q}$ and $R_{v}$. 
If the temperature is non-vanishing and exceeds 
$\hbar \omega$, Eq. (\ref{effectiveR}) holds with $\hbar \omega$
replaced by $kT$. 
Below we discuss the resistances $R_{q}$ and $R_{v}$ in detail for 
several examples: a chaotic cavity, ballistic wires, 
a Hall bar and a quantum point contact. 

\subsection{Chaotic Cavities}

\begin{figure}
\epsfysize=9cm
\epsfxsize=7cm
\centerline{\epsffile{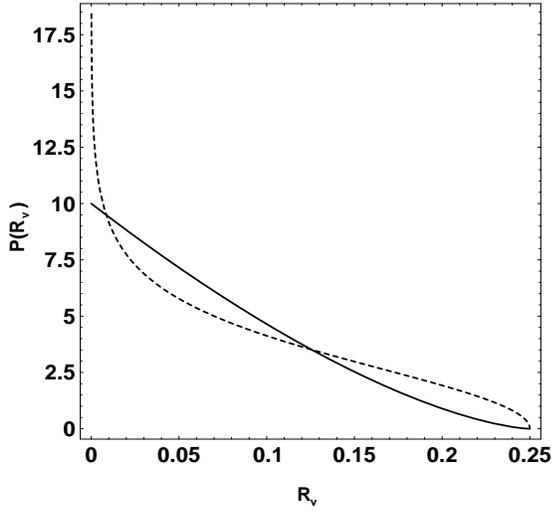}}
\vspace*{0.3cm}
\caption{ \label{qdrv}
Distribution of the resistance $R_v$ of a chaotic quantum dot
for the orthogonal ensemble (dashed) and the unitary ensemble (solid line) 
line. After Ref. \protect\cite{pedersen}.}
\end{figure}

Since the equilibrium charge relaxation resistance is the 
sum of equally weighted diagonal density matrix elements 
the distribution of the charge relaxation resistance 
shown in Fig.~\ref{qdrq} and given by Eq. (\ref{distr})
for the mesoscopic capacitor coupled to a single lead with 
two quantum channels is also the distribution function 
for the charge relaxation resistance of a conductor 
coupled to two leads each supporting one quantum channel. 
Novel information is, however, obtained if we consider 
the resistance $R_v$ which governs the charge fluctuations
in the presence of a dc-current through the cavity. 
For the resistance $R_v$ 
the distribution\cite{pedersen} is shown in Fig.~\ref{qdrv}.
It is limited to the range
$0, h/4e^{2}$ and (in units of $h/e^2$) given by 
\begin{equation}
P(R_v) = \left\{ 
\begin{array}{ll}
2 \log \left[\frac{1-2R_v+\sqrt{1-4R_v}}{2R_v}\right], & \beta=1, \\
10 (1-4 R_v)^{3/2}, & \beta=2.
\end{array}
\right.
\end{equation}
For the orthogonal ensemble the distribution is singular at $R_v = 0$.
Both distribution functions tend to zero at $R_v = 1/4$.

The charge relaxation resistance of a chaotic cavity coupled 
to two reservoirs\cite{brouwer1}
via two point contacts with $N_{1}>>1$ and $N_{2}>>1$ open quantum channels
is given by Eq. (\ref{RqN}) with $N = N_{1} + N_{2}$. 
It is illustrative to compare the charge relaxation resistance 
Eq. (\ref{RqN}) with the dc-resistance of this cavity 
which after ensemble averaging is given by 
\begin{eqnarray}
    G = \frac{h}{e^2}\left(\frac{N_{1}N_{2}}{N} 
    - \frac{2-\beta}{\beta} \frac{N_{1}N_{2}}{N^{2}} + O(N^{-3}) \right) .
	\label{gc} 
\end{eqnarray}
Here the first term is the classical resistance and the second term is 
a small weak localization correction. 
The average conductance is obtained by adding the resistances 
of the quantum point contact in {\it series} $G^{-1} = ({e^2}/h) (1/N_1 +1/N_2)$. 
In contrast the average charge relaxation resistance is 
obtained by adding the resistances of the quantum point 
contacts in {\it parallel} and by observing that each channel contributes 
with a resistance quantum of $2h/e^{2}$, 
$R_q^{-1} = (2{e^2}/h) (N_1 + N_2)$. 
For the case that both contacts support an equal number of channels
the two resistances differ just by a factor of eight. 
On the other hand in the case that the two contacts are very different 
$N_{1} >>N_{2}$ the dc-conductance is essentially determined by the 
smaller quantum point contact whereas the charge relaxation resistance 
is essentially determined by the large point contact. 

\subsection{Ballistic Wire, Edge States} 

\begin{figure}
\narrowtext
{\epsfxsize=7cm\epsfysize=3.0cm\centerline{\epsfbox{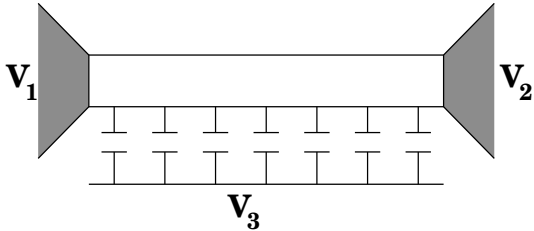}}}
\vspace*{0.3cm}
\caption{The 1D wire, connected to two reservoirs and coupled 
capacitively to a gate. After Ref. \protect\cite{bhb}. }
\label{fig1}
\end{figure}

So far we have considered zero-dimensional systems 
in which the internal potential $U$ is described by a single variable. 
In reality we always deal with a potential landscape. The potential 
is in general a complicated function of position. We thus expect that 
in general the charge relaxation reflects the potential landscape
and thus depends more directly on the long range Coulomb interaction than 
indicated by 
the simple formulas given above. Interestingly, as long as
the ground state can be taken to be uniform, the charge relaxation 
resistance turns out to be independent of the Coulomb interaction even in
spatially extended systems. 
Consider for instance the long one-channel wire shown in Fig.~\ref{fig1}.
The wire is coupled to two electron reservoirs which serve as carrier 
sources and sinks and is in close proximity 
to a gate. If $c$ is the 
geometric capacitance 
of the wire to the gate per unit length, the Coulomb interaction adds 
an energy 
\begin{eqnarray}
\int dx \frac{e^{2}}{2c} \rho^{2} (x)
\label{ecoul} 
\end{eqnarray}
where $\rho$ is the excess density of electrons on the wire. 
Such a potential energy term is characteristic for Luttinger models
for which the interaction term is written as $\nu^{-1}\int dx (1-1/g^{2})
\rho^{2} (x)$ where $g$ is the interaction parameter and 
$\nu = 1/hv$ the density of states at the Fermi 
energy. Here $v$ is the Fermi velocity.  
Comparison of 
the two terms gives a connection between the geometrical 
capacitance, the density of states and the interaction parameter, 
$g^{2} = 1/(1+ e^{2}\nu/c)$. As a consequence the electrochemical 
capacitance of the wire to the gate $c_{33} = c_\mu$ can be found 
by using Eq. (\ref{c0}) and per unit length can be written as\cite{bhb}
\begin{eqnarray}
c_{\mu} = g^{2} e^{2}\nu . 
\end{eqnarray}
If the variation of the potential near the contacts is neglected 
the ground state can be taken to be uniform. The potential 
which develops if one of the contact voltages is varied 
is non-uniform and depends on the frequency. 
Nevertheless, Blanter et al. \cite{bhb}
find that 
a one-channel wire of length $L$, coupled capacitively to a gate, has a low 
frequency expansion of $G_{33} \equiv G (\omega)$ given by 
\begin{eqnarray}
G (\omega)  = -i\omega C_{\mu} + \omega^{2} C_{\mu}^{2} R_{q}+O(\omega^{3})
\label{gbal} 
\end{eqnarray}
with $C_{\mu} = Lc_{\mu}$ and a charge 
relaxation resistance given by 
\begin{eqnarray}
R_{q} = h/4e^{2}. 
\label{Rqbal} 
\end{eqnarray}
Note again that the charge relaxation resistance corresponds to the 
parallel addition of two conductances (one per contact) of 
$2e^{2}/h$. It is shown in Ref. \cite{bhb} that a measurement of 
$G_{33} \equiv G (\omega)$ up to the third order in frequency permits 
to determine the interaction constant $g$. 

The quantization of the charge relaxation resistance is lost as soon 
as we consider a many channel ballistic wire. If the velocity 
of channel $i$ at the Fermi energy is $v_i$ the charge relaxation resistance 
of an N channel ballistic wire is given by\cite{bc1,bb} 
\begin{eqnarray}
R_{q} = \frac{h}{4e^{2}} \frac{\sum_{i=1}^{i=N} (\frac{1}{hv_i})^{2}}
{[\sum_{i=1}^{i=N} \frac{1}{hv_i}]^{2}}
\label{RqbalN} 
\end{eqnarray}
This resistance varies between $R_{q} = h/4e^{2} N$ when all 
density of states are comparable and $R_{q} = h/4e^{2}$ when one of the 
density of states is very much larger than all others. 

In a two dimensional electron gas in a high magnetic field
the extended states at the Fermi surface which give rise to 
the quantized Hall resistance are edge states. For each 
Landau level that is below the Fermi energy in the bulk of a Hall 
bar there exists an edge state with velocity $v^{u}_{i}$ at the 
upper edge and with velocity $v^{l}_{i}$ at the lower edge. In 
reality the edge state follows an equipotential line and the velocity 
is thus space dependent. We need the density of states $1/hv$
rather than the velocities themselves. To take the spatial variation
into account we can introduce the density of states averaged 
along an edge state over the entire length of the Hall bar. 
Furthermore, the electrostatic potential can differ for different
edge states on the same edge of the sample. Here we assume for simplicity
that all edge states 
of the sample are at the same potential $U^{u}$ 
at the upper edge and at the same potential $U^{l}$ at the lower edge.  
A dipole is established in this sample by charging the upper edge 
states against the edge states on the lower edge. The relaxation 
of the excess charge on the edge states is then determined by a
charge relaxation resistance which according to Christen and the 
author\cite{bc2} is given by 
\begin{eqnarray}
R_{q} = \frac{h}{2e^{2}} 
\left(\frac{\sum_{i=1}^{i=N} (\frac{1}{hv^{u}_i})^{2}}
{[\sum_{i=1}^{i=N} \frac{1}{hv^{u}_i}]^{2}}
+ \frac{\sum_{i=1}^{i=N} (\frac{1}{hv^{l}_i})^{2}}
{[\sum_{i=1}^{i=N} \frac{1}{hv^{l}_i}]^{2}}\right) .
\label{Rqbqhe} 
\end{eqnarray}
Like the charge relaxation resistance 
of a ballistic wire, it is dominated by 
the channel with the highest density of states (smallest Fermi
velocity). Here we can imagine that the upper edge presents 
a much smoother potential then the lower edge. In that 
case the charge relaxation resistance might be as high as 
$R_q = h/2e^{2}$. 

\subsection{Charge Relaxation Resistance of a Quantum Point Contact}

A quantum point contact is a constriction in a two
dimensional electron gas with a lateral width of the order
of a Fermi wavelength \cite{vanwees,wharam} .
We combine the capacitances 
of the conduction channel to the two gates and describe the 
interaction of the quantum point contact with the gates with the help
of a single geometrical capacitance $C$. 
A simple model of a quantum point contact takes the potential 
to be a saddle point at the center of the constriction\cite{butqpc,landm}, 
\begin{equation}
    V(x,y) = V_0 + \frac{1}{2} m \omega_y^2 y^2
    - \frac{1}{2} m \omega_x^2 x^2
\end{equation}
where $V_0$ is the electrostatic potential at the saddle 
and the curvatures of
the potential are parametrized by $\omega_x$ and $\omega_y$.
For this model the scattering matrix is diagonal, 
i.e.\ for each quantum channel ( energy $\hbar \omega_y (n+1/2)$
for transverse motion)
it can be represented as a $2\times 2$-matrix.
For a symmetric scattering potential and without a magnetic field
the scattering matrix is of the form
\begin{equation}
    s_{n}(E) = \left( \begin{array}{ll}
	-i \sqrt{R_{n}} \exp(i\phi_{n}) & \sqrt{T_{n}} \exp(i\phi_{n}) \\
	\sqrt{T_{n}} \exp(i\phi_{n}) & -i\sqrt{R_{n}} \exp(i\phi_{n})
	\end{array} \right)
\end{equation} 
where $T_{n}$ and $R_{n}= 1-T_{n}$
are the transmission and reflection probabilities 
of the n-th quantum channel and $\phi_{n}$ is the phase accumulated by a 
carrier
in the n-th channel
during transmission through the QPC. 
The probabilities for transmission through the saddle point are\cite{butqpc} 
\begin{eqnarray}
    T_{n}(E) &=& \frac{1}{1+e^{-\pi\epsilon_n(E)}} ,\\
    \epsilon_n(E) &=& 2\left[ E-\hbar\omega_y(n+\frac{1}{2})-V_0\right]/
	(\hbar\omega_x) .
\end{eqnarray}
The transmission probabilities determine the conductance 
$G = ({e^2}/{h}) \sum_{n} T_{n}$ and the zero-frequency
shot-noise \cite{buttiker90,lesovik,buttiker91,martin}
$S(\omega = 0, V) = ({e^2}/{h}) (\sum_{n} T_{n} R_{n}) e|V|$. 
As a function of energy (gate voltage) the conductance rises 
step-like\cite{vanwees,wharam}. The shot noise 
is a periodic function of energy. The oscillations 
in the shot noise associated with the opening of a quantum channel 
have recently been demonstrated experimentally in a clear and 
unam-
\begin{figure}
\narrowtext
\epsfysize=8cm
\epsfxsize=7cm
\centerline{\epsffile{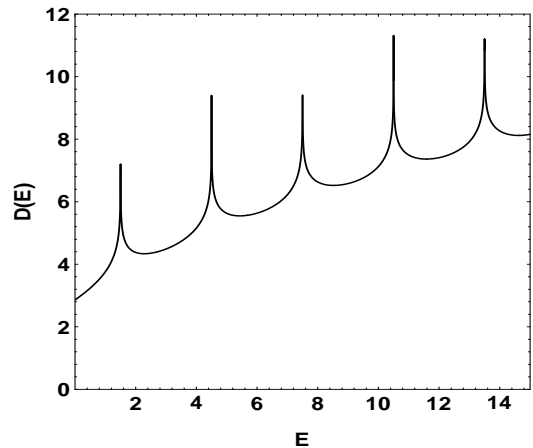}}
\vspace*{0.3cm}
\caption{ \label{dos_fig}
Density of states in units of $4/(h\omega_x)$ for a saddle-point constriction 
as function of energy, $E/(\hbar\omega_x)$. After Ref. \protect\cite{pedersen}.
}
\end{figure}
biguous manner 
by Reznikov et al.\cite{reznikov} and Kumar et al.\cite{kumar}.
Deviations from the equilibrium charge distribution 
can occur in various ways\cite{cbprl,cura}. If the capacitances to the gate are 
large the charge distribution takes the form of a dipole\cite{cbprl,aron} 
across the QPC. If the geometrical capacitance for this dipole is 
large the QPC can be charged vis-a-vis the gate\cite{cura}. 
Here we consider the latter situation. 

First we now need the density of states of the QPC. To this end 
we use the relation between 
density and phase $D_{n} = (1/\pi) \phi_{n}$
and evaluate it semi-classically.
The spatial region of interest for which we have to find 
the density of states is the region over which the electron
density in the contact is not screened completely. 
We denote this length by $\lambda$. 
The density of states is then found from  
$N_{n}=1/h \int_{-\lambda}^{\lambda} \frac{dp_n}{dE} dx$ 
where $p_{n}$ is the classically allowed momentum.
A simple calculation gives a density of states 
\begin{equation}
 {\cal   D}_{n} (E) = \frac{4}{h\omega_x}
    \mbox{asinh} \left( \sqrt{\frac{1}{2} \frac{m\omega_x^2}{E-E_n}}\lambda  \right),
\end{equation}
for energies $E$ exceeding the channel threshold $E_n$ and 
\begin{equation}
 {\cal   D}_{n} (E) = \frac{4}{h\omega_x}	    
   \mbox{acosh} \left( \sqrt{\frac{1}{2} \frac{m\omega_x^2}{E_n-E}}\lambda \right) ,
\end{equation}
for energies in the 
interval $E_n - (1/2) m \omega_x^2 \lambda^2 \leq E < E_n$ 
below the channel threshold. 
Electrons with energies less than $E_n - \frac{1}{2} m \omega_x^2 \lambda^2 $
are reflected before
reaching the region of interest, and thus do not contribute 
to the density of states.
The resulting density of states has a logarithmic singularity
at the threshold $E_{n}= \hbar\omega_y(n+\frac{1}{2})+V_0$  
of the n-th quantum channel. (We expect that a fully quantum mechanical 
calculation gives a density of states which exhibits also
a peak at the threshold but which is not singular). 
The total density of states ${\cal D} = \sum_n{\cal D}_{n}$ as function 
of energy (gate voltage) is shown in Fig.~\ref{dos_fig} 
for 
$\omega_y/\omega_x=3$, $V_0=0$ and $m\omega_x \lambda^2/\hbar=18$.
Each peak in the density of states of Fig.~\ref{dos_fig} 
marks the opening of a new channel. With the help of 
the density of states we also obtain the capacitance 
$C_\mu^{-1} = C^{-1} + ( e^2 {\cal D})^{-1}$. 

It is instructive to evaluate the resistances 
$R_q$ explicitly in terms of the parameters 
which determine the scattering matrix. 
We find for the density of states matrix 
of the n-th quantum channel,
\begin{figure}
\narrowtext
\epsfysize=8cm
\epsfxsize=7cm
\centerline{\epsffile{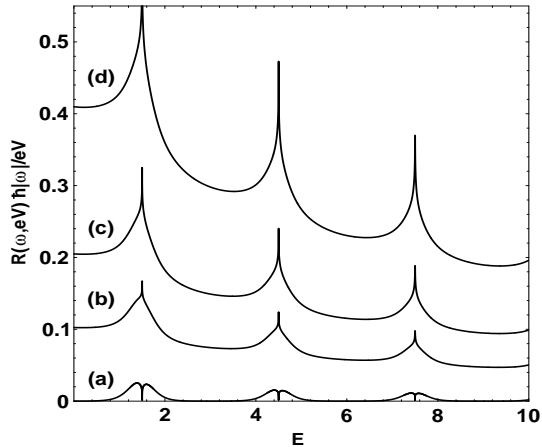}}
\vspace*{0.3cm}
\caption{ \label{R_fig}
Effective resistance, in units of $h/e^2$, as function of
energy, $E/(\hbar\omega_x)$ for the cases
$a)\hspace{0.15cm} \hbar\omega/(eV)=0$, 
$b)\hspace{0.15cm} \hbar\omega/(eV)=0.25$, 
$c)\hspace{0.15cm} \hbar\omega/(eV)=0.5$ and 
$d)\hspace{0.15cm} \hbar\omega/(eV)=1$, 
where $V$ is the bias voltage. After Ref. \protect\cite{pedersen}.}
\end{figure}
\begin{eqnarray}
    {\cal N}_{11} &=& {\cal N}_{22} = \frac{1}{2\pi} \frac{d\phi_{n}}{dE},\\
    {\cal N}_{12} &=& {\cal N}_{21} = \frac{1}{4\pi} 
    \frac{1}{\sqrt{R_{n}T_{n}}} \frac{dT_{n}}{dE}.
\end{eqnarray}
Inserting these results into 
the expression for the charge relaxation 
resistance Eq.~(\ref{Rq}) 
gives curve (d) in Fig.~\ref{R_fig}.

The resistance $R_v$ is given by \cite{pedersen} 
\begin{eqnarray}
R_v &=& \frac{h}{e^2} \frac{ \sum_n \frac{1}{4R_nT_n}
	\left( \frac{dT_n}{dE} \right)^2}{[\sum_{n} (d\phi_{n}/dE)]^{2}}.
	\label{rvqpc}
\end{eqnarray}
It is sensitive to the variation with energy of the transmission probability. 
Note that the transmission probability has the form of a Fermi function.
Consequently, the derivative of the transmission probability is 
also proportional to $T_n R_{n}$. The numerator of Eq.~(\ref{rvqpc})
is thus also maximal at the onset of a new channel and vanishes on 
a conductance plateau.

In Fig.~\ref{R_fig} the effective resistance $R(\omega,V)$ is shown 
for four frequencies $\hbar\omega/(eV)=0,0.25,0.5,1$, 
where $V$ is the applied voltage. 
At the highest frequency  $\hbar\omega/(eV)= 1$ 
the resistance $R(\omega,V)$ is completely dominated by the equilibrium 
charge relaxation resistance $R_q$. 
The uppermost curve (d) of Fig.~\ref{R_fig} is nothing but $R_q$ and 
determines the noise due to the zero-point equilibrium fluctuations. 
The fluctuations reach a maximum at the onset of 
a new channel since $R_q$ takes its maximum value, $R_q = h/e^{2}$.
At the lowest frequency $\hbar\omega= 0$ the resistance $R(\omega,V)$
is determined by $R_v$. 
The lowermost curve (a) of Fig.~\ref{R_fig} is the nonequilibrium 
resistance $R_v$. 
It is seen that the non-equilibrium resistance $R_v$ is
very much smaller than $R_q$. We have encountered such 
a difference between $R_q$ and $R_v$ already for the chaotic cavities. 
Furthermore $R_v$ exhibits a double peak 
structure: The large peak in the density of states at the threshold 
of a quantum channel nearly suppresses the non-equilibrium noise 
at the channel threshold completely. Two additional 
curves (b and c for $\hbar\omega/(eV)= 0.25$ and  $\hbar\omega/(eV)= 0.5$) describe 
the crossover from $R_v$ to $R_q$. 

The resistances $R_q$ and $R_v$ probe an aspect of mesoscopic conductors
which is not accessible by investigating the dc-conductance or the 
zero-frequency limit of shot noise. 
A more realistic treatment of the charge distribution
of a quantum point contact includes a dipole across the quantum point 
contact itself\cite{cbprl,aron} and in the presence of the gates 
includes a quadrupolar charge distribution\cite{cura}.

\section{Discussion} \label{conclusion}

In this work we have summarized our current 
understanding of the relaxation 
of excess charge and its fluctuations both at equilibrium and 
in a transport state of mesoscopic conductors. 
At low frequencies the dynamics of the 
charge is governed by an out-of-phase response which is given 
by an electrochemical capacitance or a kinetic inductance. 
The dissipative, in-phase response is determined by a charge relaxation resistance.
For the simple models investigated here, we find that the charge relaxation 
resistance is universal in the single channel limit and 
in the two channel limit is universal for systems with a uniform 
ground state (ballistic wire). 
In general, for systems connected to two or more channels 
the charge relaxation resistance is sample specific. 
For the case of chaotic cavities connected to two 
channels we have discussed the distribution of the charge relaxation 
resistances. For cavities connected to many channels the average 
charge relaxation resistance has a classical part and a quantum mechanical
weak localization correction. 
We have also discussed the charge relaxation resistance 
for systems subject to quantizing magnetic fields
and found that it is dominated by the edge channel with the 
highest density of states. For a quantum point contact we have found 
that the charge relaxation resistance peaks at the opening of a new channel.

While we have discussed charge relaxation resistances for 
a number of different conductors, there are still a number of 
important cases which we have not treated. For instance, 
the charge relaxation resistance of a metallic diffusive 
mesoscopic wire is 
not known \cite{texier}.  For a metallic wire close to a gate, we expect 
that a small variation of the gate voltage changes the potential 
only near the surface of the metallic wire within a depth of a 
screening length. Thus the charge relaxation resistance can be expected to 
be determined by the time it takes the excess charge 
which resides in this screening layer to move in and out of the conductor. 
The considerations presented above for the 
quantized Hall conductors should be extended to the case 
where we are in between plateaus and the bulk states becomes 
important. This problem thus requires also an understanding 
of charge relaxation in insulators. 

We have also investigated the fluctuations of the charge. 
At equilibrium these fluctuations are governed by the 
charge relaxation resistance. The charge fluctuations 
in a transport state are governed by a non-equilibrium charge relaxation 
resistance. We expect that the internal potential fluctuations are 
important for the discussion of dephasing in small conductors. 

The charge relaxation of mesoscopic conductors
is clearly a very interesting subject with so far little 
theoretical and experimental interest. We hope that this work 
stimulates a wider interest for the charge dynamics of mesoscopic
conductors.

\acknowledgments

This work was supported by the Swiss National Science
Foundation. 



\end{multicols}


\begin{references}
%
\bibitem{btp93} M.~B\"uttiker, H.~Thomas, and A.~Pr\^etre,
                Phys.\ Lett.\ A {\bf 180}, 364 (1993).
%
\bibitem{cura}  M.\ B\"{u}ttiker and T.\ Christen, in {\em Mesoscopic Electron
                Transport}, ed.\ by L.~P.~Kouwenhoven, G.~Sch\"on, 
                and L.~L.~Sohn, NATO
                ASI Series E (Kluwer Academic Publishing, Dordrecht, 
                1997). Vol. 345. p. 259.
                
%
\bibitem{cbprl} T.\ Christen and M.\ B\"uttiker, Phys.\ Rev.\ Lett.\,
                {\bf 77}, 143 (1996). 
                
\bibitem{cbepl} T. Christen and M. B\"uttiker, 
                Europhys. Lett. {\bf 35}, 523 (1996).
                
\bibitem{song}  A. M. Song, A. Lorke, A. Kriele, J. P. Kotthaus, 
                W. Wegschneider, M. Bichler, Phys. Rev. Lett. {\bf 80}, 
                3831 (1998). 

\bibitem{ma}    Z. S. Ma, J. Wang, and H. Guo, Phys. Rev. B {\bf 57}, 9108
                (1998). 
                               
\bibitem{sablikov}  V. A. Sablikov and B. S. Shschamkhalova, Phys. Rev. 
                    {\bf 58}, 13847 (1998). 
                                                            
\bibitem{anat}  M. P. Anatram, J. Phys. Cond. Mat. {\bf 10}, 9015 (1998). 

\bibitem{aron}  I. E. Aronov, et al. Phys. Rev. B {\bf 58}, 9894 (1998). 

\bibitem{kuro}  S. Kurokawa and A. Sakai, J. of Applied Physics, {\bf 83}, 
                7416 (1998). 
                
\bibitem{photo} M. H. Pedersen and M. B\"uttiker, Phys. Rev. B {\bf58}, 
                12993 (1998).  
                             
\bibitem{agua}  R. Aguado and G. Platero, Phys. Rev. Lett. {\bf 81}
                4971 (1998). 
                
\bibitem{tang} C. S. Tang and C. S. Chu, (unpublished). 

\bibitem{smith} F. T. Smith, Phys.\ Rev.\ {\bf 118} 349 (1960).
%
\bibitem{mb83}  It is the derivative with respect to the potential and 
                not the energy which are fundamental. The phase derivatives 
                with respect to the potential lead to a dwell time which 
                differs from the Wigner-Smith phase-delay times,
                M. B\"{u}ttiker, Phys. Rev. B {\bf 27}, 6178 (1983) and 
                C. R. Leavens and G. C. Aers, Phys. Rev. B{\bf 39}, 1202
                (1989). 
%
\bibitem{mb90}     M. B\"{u}ttiker, in "Electronic Properties of Multilayers 
                   and low
                   Dimensional Semiconductors",
                   edited by J. M. Chamberlain, L. Eaves, and J. C. Portal,
                   (Plenum, New York, 1990). p. 297.
                
\bibitem{imry}     Y. Imry, in {\it Directions in Condensed Matter Physics},
                   edited by G. Grinstein  and G. Mazenko, 
                   (World Scientific Singapore, 1986). p. 101. 

\bibitem{land}    R. Landauer, Z. Phys. B {\bf 68}, 217 (1987).  

\bibitem{butt89}  M.\ B\"uttiker, Phys.\ Rev.\ Lett.\ {\bf 57}, 1761 (1986);
                  Phys.\ Rev.\ B {\bf 40}, 3409 (1989).

\bibitem{gram}    T.\ Gramespacher and M.\ B\"uttiker, 
                  Phys.\ Rev.\ B {\bf 56}, 13026 (1997).
                  
\bibitem{sols}    F. Sols and J. Sanchez-Cañizares, cond-mat/9811325 


\bibitem{mbjmp}   M. B\"uttiker, J. Math. Phys., {\bf 37}, 4793 (1996).

\bibitem{aleiner} A. Kaminski, I. L. Aleiner, L. I. Glazman, 
                  Phys. Rev. Lett. {\bf 81}, 685 (1998); 
                  I. L. Aleiner and L. I. Glazman,
                  Phys. Rev. B {\bf 57}, 9608 (1998). 

\bibitem{marcus}  C. M. Marcus, R. M. Westervelt, P. F. Hopkins and 
                  A. C. Gossard, 
                  Phys. Rev. B {\bf 48}, 2460 (1993);  
                  M. Switkes, C. M. Marcus , K. Campman, A. C. Gossard, 
                  (unpublished). 

\bibitem{beenakker} C.~W.~J.\ Beenakker, 
                    Rev.\ Mod.\ Phys.\ {\bf 69}, 731 (1997).

\bibitem{fyod2}     Y.~V.\ Fyodorov and H.~J.\ Sommers, 
                    Phys. Rev. Lett. {\bf 76}, 4709 (1996). 

\bibitem{gopar1}    V.~A.\ Gopar, P.~A.\ Mello, and M.\
                    B\"{u}ttiker, Phys.\ Rev.\ Lett.\ {\bf 77}, 3005 (1996).  
 
\bibitem{gopar2}    V. A. Gopar and P. A. Mello, Europhys. Letters, 
                    {\bf 42}, 131 (1998).

\bibitem{brouwer2}  P.~W.\ Brouwer, K.~M.\ Frahm, and
                    C.~W.~J.\ Beenakker, Phys.\ Rev.\ Lett.\ {\bf 78}, 
                    4737 (1997).

\bibitem{brouwer3}  P.~W.\ Brouwer, S.~A.\ van Langen, K.~M.\ Frahm, 
                    M.\ B\"uttiker and C.~W.~J.\ Beenakker, 
                    Phys.\ Rev.\ Lett.\, (1997).

\bibitem{stijn98}   S. F. Godijn, S. Moeller, H. Buhmann, L. W. Molenkamp, 
                    S. A. van Langen,  cond-mat/9811181.
     

\bibitem{pedersen}  M. H. Pedersen, S. A. van Langen, M. B\"{u}ttiker,
                    Phys. Rev. {\bf B 57}, 1838 (1998).
                    
\bibitem{genack}    P. Sebbah, O. Legrand, A. Z. Genack,cond-mat/9808237 .

\bibitem{brouwer4}  P. W. Brouwer and C. W. J. Beenakker, 
                    J. Math. Phys. {\bf 37}, 4904 (1996). 
                   
\bibitem{brouwer1}  P.~W.\ Brouwer and M.\ B\"uttiker, 
                    Europhys.\ Lett.\ {\bf 37}, 441 (1997).

\bibitem{buttiker90} M.\ B\"uttiker, Phys.\ Rev.\ Lett.\, {\bf 65}, 
                     2910 (1990); Phys.\ Rev.\ B {\bf 46}, 12485 (1992).

\bibitem{bhb}       Ya. M. Blanter, F.W.J. Hekking, and M. B\"uttiker, 
                    Phys. Rev. Lett. {\bf 81}, 1749 (1998).              
 
\bibitem{bc1}       M. B\"{u}ttiker and T. Christen, in 
                    {\it Quantum Transport in Semiconductor Submicron Structures},
                    edited by B. Kramer, (Kluwer Academic Publishers, Dordrecht, 1996);
                    NATO ASI Series, Vol. {\bf 326}, 263 (1996).

\bibitem{bb}        Y. M. Blanter and M. B\"uttiker, 
                    Europhys. Lett. {\bf 42} 535 (1998).  
                    
\bibitem{bc2}       M. B\"{u}ttiker and T. Christen, in 
                    'High Magnetic Fields in the Physics of Semiconductors', 
                    edited by G. Landwehr and W. Ossau, (World Scientific, 
                    Singapur, 1997).
                    p. 193.                   

\bibitem{vanwees}   B.~J.\ van Wees, H.\ van Houten, C.~W.~J.\ Beenakker,
                    J.~G.\ Williamson, L.~P.\ Kouwenhoven, D.\ van der Marel 
                    and C.~T.\ Foxon,
                    Phys.\ Rev.\ Lett.\ {\bf 60}, 848 (1988).

\bibitem{wharam}    D.~A.\ Wharam, T.~J.\ Thornton, R.\ Newbury, M.\ Pepper,
                    H.\ Ahmed, J.~E.~F.\ Frost, D.~G.\ Hasko, D.~C.\ Peacock, 
                    D.~A.\ Ritchie and
                    G.~A.~C.\ Jones, J.\ Phys.\ C {\bf 21}, L209 (1988).
                    
\bibitem{butqpc}    M.\ B\"{u}ttiker, Phys.\ Rev.\ B {\bf 41}, 7906 (1990).

\bibitem{landm}     A. G. Scherbakov, E. N. Bogachek, and U. Landman,
                    Phys. Rev. B {\bf 53}, 4054 (1996).
                  
\bibitem{lesovik}   G.~B.\ Lesovik, Pisma Zh.\ Eksp.\ 
                    Teor.\ Fiz.\ {\bf 49}, 513 (1989) 
                    (JETP Lett.\ {\bf 49}, 592 (1989)).
                    
\bibitem{buttiker91} M. B\"{u}ttiker,
                     Physica B {\bf 175}, 199 (1991).  

\bibitem{martin}    Th.\ Martin and R.\ Landauer, Phys.\ Rev.\ B {\bf 45},
                     1742 (1992).
                       
\bibitem{reznikov}  M.\ Reznikov, M.\ Heiblum, H.\ Shtrikman, and D.\
                    Mahalu, Phys.\ Rev.\ Lett.\ {\bf 75}, 3340 (1995).

\bibitem{kumar}     A.\ Kumar, L.\ Saminadayar, D.~C.\ Glattli,
                    Y.\ Jin, and B.\ Etienne,
                    Phys.\ Rev.\ Lett.\ {\bf 76}, 2778 (1996).  

\bibitem{texier}    For one-dimensional motion in random potentials 
                    the distribution of the global phase-delay is knwon:
                    See C. Texier and A. Comtet, cond-mat/9812196 and 
                    J. Phys. A: Math. Gen. {\bf 36}, 8017 (1997);
                    S. K. Joshi and and A. M. Jayannavar, cond-mat/9712249;
                    A. M.  Jayannavar, C. V. Vijayagovindan and N. Kumar, 
                    Z. Phys. B {\bf 75}, 77 (1989). 
                    
\end{references}
\end{document}